\documentstyle[aps,epsf,rotate,multicol]{revtex}

\begin{document}

\draft

\title{Onset of criticality and transport in a driven diffusive system}

\author{M\'aria Marko\v sov\'a$^{a,b)}$,  
	Mogens H. Jensen$^{b)}$, 
	Kent B{\ae}kgaard Lauritsen$^{b)}$, 
	and Kim Sneppen$^{c)}$
}

\address{
$^{a)}$Inst.\ of Measurement Science, Slovak Academy of Sc.,
        D\'ubravsk\'a cesta 9, 842 19 Bratislava, Slovakia
\\
$^{b)}$Niels Bohr Institute and $^{c)}$NORDITA,
	Blegdamsvej 17,
	DK-2100 Copenhagen, Denmark}

\date{Submitted: October 2, 1996; Printed: \today}

\maketitle

\begin{abstract}
We study transport properties in a 
slowly driven diffusive system where the transport 
is externally controlled by a parameter $p$. Three types of
behavior are found: For $p<p'$ the system is not conducting at all.
For intermediate $p$ a finite fraction of the
external excitations propagate through the system.
Third, in the regime $p>p_c$ the system
becomes completely conducting. For all $p>p'$
the system exhibits self-organized critical behavior.
In the middle of this regime, at $p_c$, the system undergoes
a continuous phase transition described by critical exponents.
\end{abstract}



\begin{multicols}{2}

Cascade events where
a single external perturbation may amplify in an avalanche
like dynamics are seen in a variety of controllable 
systems, such as nuclear reactors \cite{NR}, 
avalanche diodes \cite{AD} and lasers \cite{LA}.
In statistical physics one considers cascade or avalanche
events in a certain type of slowly driven diffusive systems. 
The prototype of such dynamics is the sandpile model of
Bak, Tang, and Wiesenfeld \cite{BTW}. 
This model exhibits large scale fluctuations
with avalanches similar to the examples mentioned above.
The difference being that the slowly driven systems self
organize to the critical transfer ratio, needed to see avalanches of
all sizes; models exhibiting this phenomena are called
self-organized critical (SOC).
Following this idea, a number of similar models have been proposed.
In particular, Zhang \cite{Zhang} and Manna \cite{Manna} 
proposed sandpile models
where the local redistribution contained a random element.
This defined another universality
class than the original BTW model \cite{Biham}.
Recently, similar ideas have been used in the 
explicit modelling (by \cite{CF,Amaral96-1}) of the observed 
rice pile dynamics in an experiment performed by the
Oslo group \cite{Frette96}.  In fact, the rice pile models
in \cite{CF,Amaral96-1} belong to a universality class which
encompasses a variety of different self-organized critical models,
including both earthquake models, interface depinning models \cite{PB}
and the Manna model \cite{NS}.
Discrete models in 1-d which do not fall into this class exist,
namely the deterministic models in \cite{Kadanoff}
and their stochastic variants proposed in
\cite{Lana-Kbl}.

In the present paper, we investigate 
the onset and robustness of criticality in the above
mentioned ``rice-pile'' class of models.
To this end, we introduce a control parameter
$p$ in a model which displays SOC, and investigate the robustness
of the critical state as function of this control parameter.
The control parameter will influence the so called transfer ratio, $J$,
which is defined as the
probability that an added particle will be transported through
the system. When $p$ is large the ratio will be critical,
and when it is small the process will come to an end abruptly.
The novel aspect of our study is that the model will be critical
in a whole range of values of the control parameter.
Further, in contrast to other SOC models with continuous
parameters \cite{Earth}, the critical properties
of our model does not depend on the 
control parameter as long as this is above a certain 
threshold point, $p'$. 

One may consider various formulations of
stochastic SOC models in 1-dimension. One is the
rice pile version where the particles perform a directed
walk through the system, and another is the
``slope version'' where one only considers the equivalent
redistribution of slopes (local stresses). As numerically
demonstrated in \cite {NS},
the latter is similar to the Manna model \cite{Manna}.
Here we present the model in the ``rice-pile''
language, since this allows for a direct counting
of transport properties in terms of the number of particles which
are absorbed/transported through the system.
For physical applications one may consider the added
particles as excitations. 

The model we study is defined as a simple extension
of the one introduced in \cite{Amaral96-1}.  Consider a lattice $[1,L]$ 
on which particles are added consecutively,
to the column $i=1$. The results to be reported in the following
have been obtained for $L=200$.
The state of the system is defined in terms
of the height profile $h(i)$ of the pile, or equivalently in terms
of the slopes $z_i=h(i)-h(i+1)$. 
The dynamics of the model is defined in two steps, of which 
1) only takes place when 2) has come to an end:
%
1) An avalanche is initiated by adding one 
particle to the column $i=1$.
If $z_1>1$ the column $i=1$ is considered active.
%
2) During each step in the avalanche all columns $i$ which are active 
can transfer one particle to the adjacent position $i+1$. 
The probability for each such toppling is $p$, and in case of
a toppling at position $i$, then one particle is transferred 
from column $i$ to $i+1$.
For all toppled particles, one tests whether any of the new local slopes
$z_{i-1}$, $z_i$, $z_{i+1}$ fulfills $z_j>1$.
If this is the case for a column $j$, then one particle at column 
$j$ is considered active in next step.
The procedure is then repeated with the new active sites,
until finally there are no more active sites.
Then a new particle must be added to the first column ($i=1$).

Notice that the model contains one parameter, $p$, which
describes the toppling probability for slopes $z_i>z_c \equiv 1$.
Thus, this toppling probability does not depend on the value
of the slope when it exceeds $z_c$.
As a result, the model allows for very steep slopes,
and consequently also for very large storage of 
externally imposed particles. It is this open storage capacity
which allows the model to display
both subcritical and critical behavior.

As usual, when considering systems that are slowly driven, one has to
run the dynamics for a certain transient time before
its stationary features can be studied. 
In Fig.\ 1, we show the average transport ratio $J$,
defined as the relative flux of outgoing particles to ingoing particles
in the stationary state. This is equivalent to
the probability that an added particle at position $i=1$
will escape at position $i>L$, as function of the parameter $p$.
We observe three regimes: below $p' \approx 0.53865$ 
the system is nearly completely isolating
and nothing penetrates. In the interval $[p',p_c]$, with $p_c \approx 0.7185$,
the system is partially conducting 
with a slope $z=\langle z_i \rangle = [h(1)-h(L)]/L$,
which fluctuates around a constant value determined by $p$ since
both $h(1)$ and $h(L)$ grow at the same rate in this regime.
Finally, for $p>p_c$, the system is completely conducting,
with a transport that exhibits the white noise behavior seen in Fig.\ 2. 

Before entering the detailed discussion of the  
behavior of the system around the two
transition points, $p'$ where conduction starts,
and $p_c$ where the motion of the pile gets pinned,
we first investigate the fluctuations in the dynamics inside the pile.
In Fig.\ 2 we show the fluctuations of $1-J$
as function of time for two $p$ values, $p=0.6$ and $p=0.75$.
For $p<p'$ (and large $L$), no particles ever reach $i=L$ 
(i.e., $J=0$).
For higher $p$ values, the 
transport out of the system is noisy, 
with a behavior which changes dramatically when passing $p=p_c$.
In fact (see Figs.\ 2a and 2b) we would like to
stress that although the dynamics
in both intervals $[p',p_c]$ and $[p_c,1]$ are conducting,  
the fluctuations in the transport properties
have a phase transition at $p_c$: they exhibit 
a Brownian motion below $p_c$, but a white noise signal above.

In Fig.\ 3 we plot the size of the relaxation events
as measured by integrating over the number of topplings
due to a single added particle.
We see that for $p>p'$, the probability
for having an avalanche of $s$ topplings exhibit the
power-law scaling
\begin{equation}
	p(s) \propto 1/s^{\tau}, ~~~~~~  \tau=1.57 \pm 0.05 .
\end{equation}
This result is in accordance with what we expect for the 
rice-pile universality class for 1-d systems
driven at the boundary \cite{CF,Amaral96-1,PB,NS}.
Thus, the SOC state in our system is not affected by the 
possibility that some particles permanently are absorbed in the system,
as is the case for $p'<p<p_c$. The type of nonconservation 
imposed in this interval of $p$ values does not destroy scaling. 
For $p<p'$ the situation is different and
the avalanche size distribution becomes exponentially bounded.
When no particles
are allowed to pass through the system, the system
cannot build up long-range correlations.

In Fig.\ 4, we show the system properties at respectively
the onset to conductivity ($p=p'$) and at the
depinning transition ($p=p_c$).
The onset to conductivity is the simplest to understand.
For low $p$ the chance that excitations 
propagate through the system is small.
The cascade triggered by adding one particle to the pile
at $i=1$ dies out exponentially fast.
Next, to understand the origin of the transition point 
$p=p'$ consider
the cascade process at a point $i>1$.
A particle toppling from position $i$ makes particles 
at $i-1$, $i$ and $i+1$
potentially active. If the corresponding slopes take values 
larger than $z_c (\equiv 1)$, 
each of the columns will topple with probability $p$ and make
the new columns active. In particular, when part of the pile
is in a state with slopes $z_j \gg 1$ then the 
toppling events will always generate new 
active columns which topple exactly with probability $p$.
This means that the spreading of active particles are determined
by directed percolation (DP) on a square lattice with $3$ descendents.
This process has a critical point at $p=0.53865$
which is in complete agreement with our numerical estimate
for $p'$, thus we use $p'=0.53865$.
Below the value $p'$ the spreading of activity dies out and
the $z_i$ values increases indefinitely, making our subcritical DP
picture self consistent. 

For $p>p'$, the DP becomes supercritical, and the system
becomes at least partially
conducting. Thus the local slopes do not grow indefinitely.
Accordingly, the DP mapping breaks down,
and the spreading of activity in the lattice will also
be influenced by dynamically distributed absorbing states
(states with slope $0$ and $1$) in the system.
The transition at $p=p'$ may be quantified as in Fig.\ 4a, 
where we show
the vanishing of the current as function of the distance to $p'$.
In order to obtain a better scaling we subtract the finite current $J'$
at $p'$,
which occur only because we have a finite system.
Given that $p'=0.53865$ is exactly known from DP studies, and measuring
the finite current for a system of size 
$L=200$ at $p=p'$ to $J' \equiv J(0.53865)=0.013$,
we observe numerically that $J(p)-J' \propto (p-p')^{\delta'}$
with $\delta' \approx 0.9$. 

Approaching $p=p_c$, the system develops an ever 
increasing region extending from $i=1$ 
where many of the local slopes are small.
At $p=p_c$, this region spans the entire system,
and all particles added to $i=1$ will eventually be conducted out of
the system at $L$ (see Fig.\ 1). Above $p_c$ the 
system profile remains bounded (pinned) at position $i=L$.
By decreasing $p$, the profile starts moving upwards 
with a velocity
that depends on $p_c-p$. In Fig.\ 4b, we see that the 
number of particles absorbed in the system per unit time scales as
\begin{equation}
	1-J \propto |p_c-p|^{\delta}, ~~~~~~~ \delta=0.9 \pm 0.1  .
\end{equation}
In the language of rice pile 
dynamics this would be the ``velocity'' of the pile $(=h(L)/t)$, 
and accordingly the transition at $p=p_c$ may be understood as a depinning
transition for the profile of the rice pile.
For $p$ values on both side of this transition
the system is critical, with exponents determined by 
the ``Manna'' universality class (with avalanche exponent $\tau=1.55$
and avalanche dimension $D=2.25\pm0.05$).
It is an open question whether the depinning exponent 
$\delta$ can be related to the universal exponents of 
this very robust class of SOC models.
\vspace{0.5cm}

In summary, we have introduced and analysed a 
slowly driven diffusive system which exhibits
two phase transitions: the first described by the critical point of
directed percolation, and it defines a transition
to a self-organized critical state described by a very robust
universality class.
The second transition, at $p_c$,  happens within the SOC state 
and can best be characterized as
a transition to a state where the
long-range meandering of the transport properties 
seen below $p_c$ gets pinned at all values above $p_c$.
We find it interesting that conductivity and the presence of a SOC
state is intimately related in the present modelling. 
\vspace{0.5cm}

M.M. thanks for the financial support of the PECO-network
grant of the European community, European Project CIPD-CT94-0011,
and for the support of the Slovak Academy of Sciences, grant 2/2025/96.
K.B.L. thanks the Danish Natural Science Research
Council for financial support.



\begin{figure}
\narrowtext
\centerline{
	\epsfysize=\columnwidth{\rotate[r]{\epsfbox{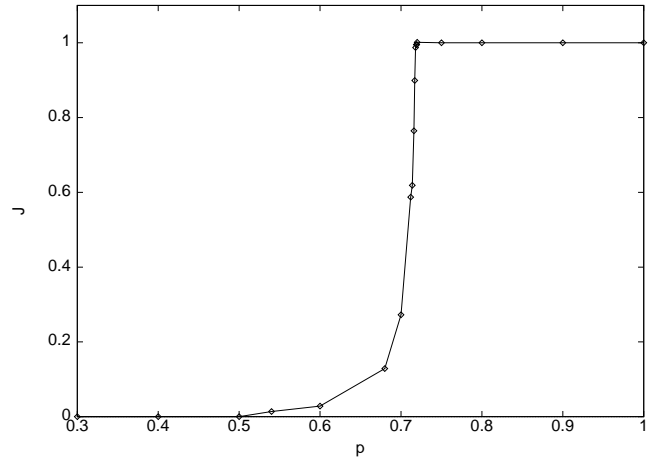}}}
}
\vspace*{0.5cm}
\caption{
Average transport ratio $J$ of particles through the system,
as function of the parameter $p$.
Transport is defined in terms of fraction of conducted particles.
Notice that when the average transport approaches unity, then in
the rice-pile picture the growth velocity 
($=1-J$) of the pile approaches zero. The point where the transport starts
is $p' \approx 0.53865$, and the point where it reaches unity
is $p_c \approx 0.7185$. 
\label{transport}}
\end{figure}

\begin{figure}
\narrowtext
\centerline{
	\epsfysize=\columnwidth{\rotate[r]{\epsfbox{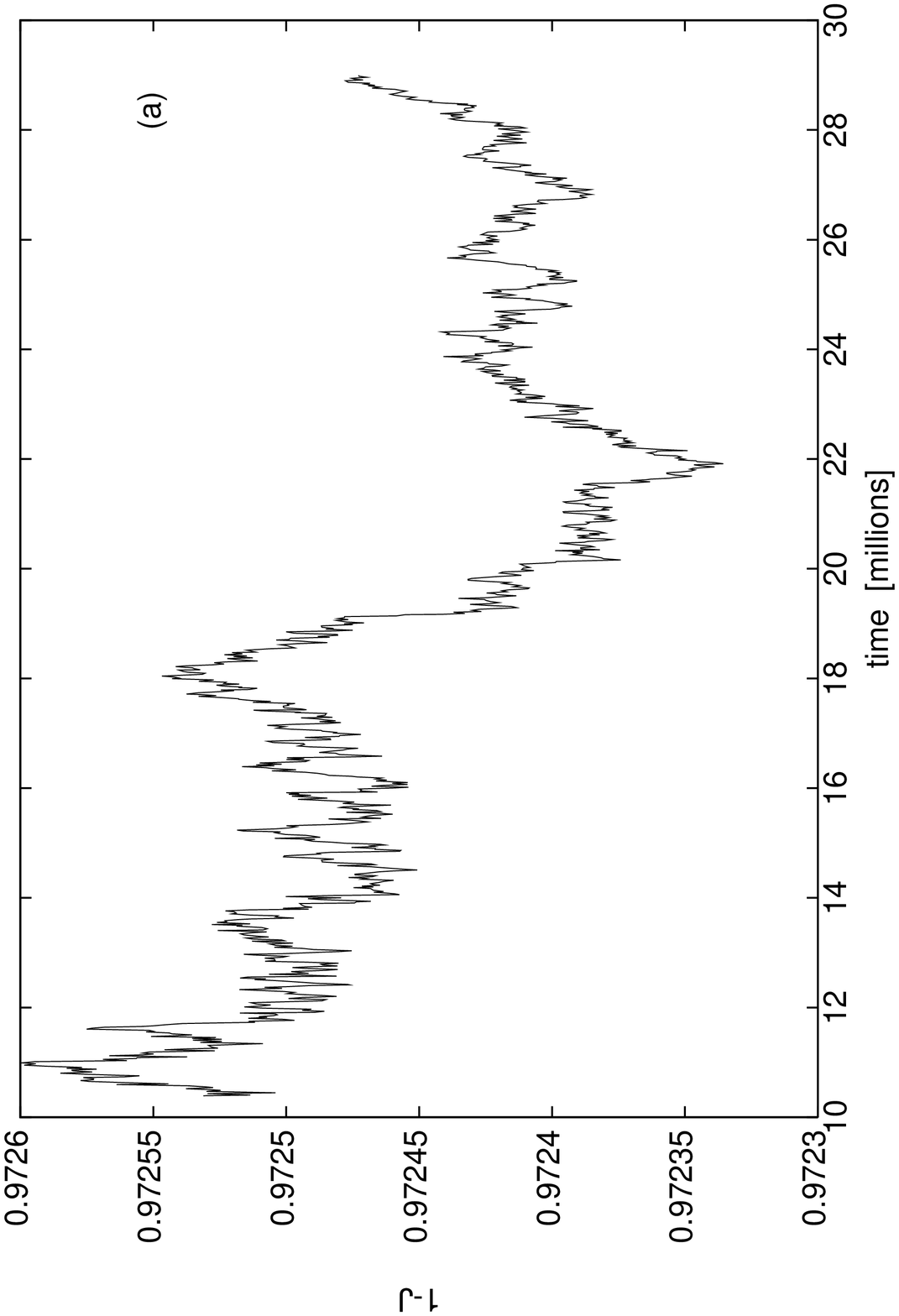}}}
}
\vspace*{0.5cm}
\centerline{
	\epsfysize=\columnwidth{\rotate[r]{\epsfbox{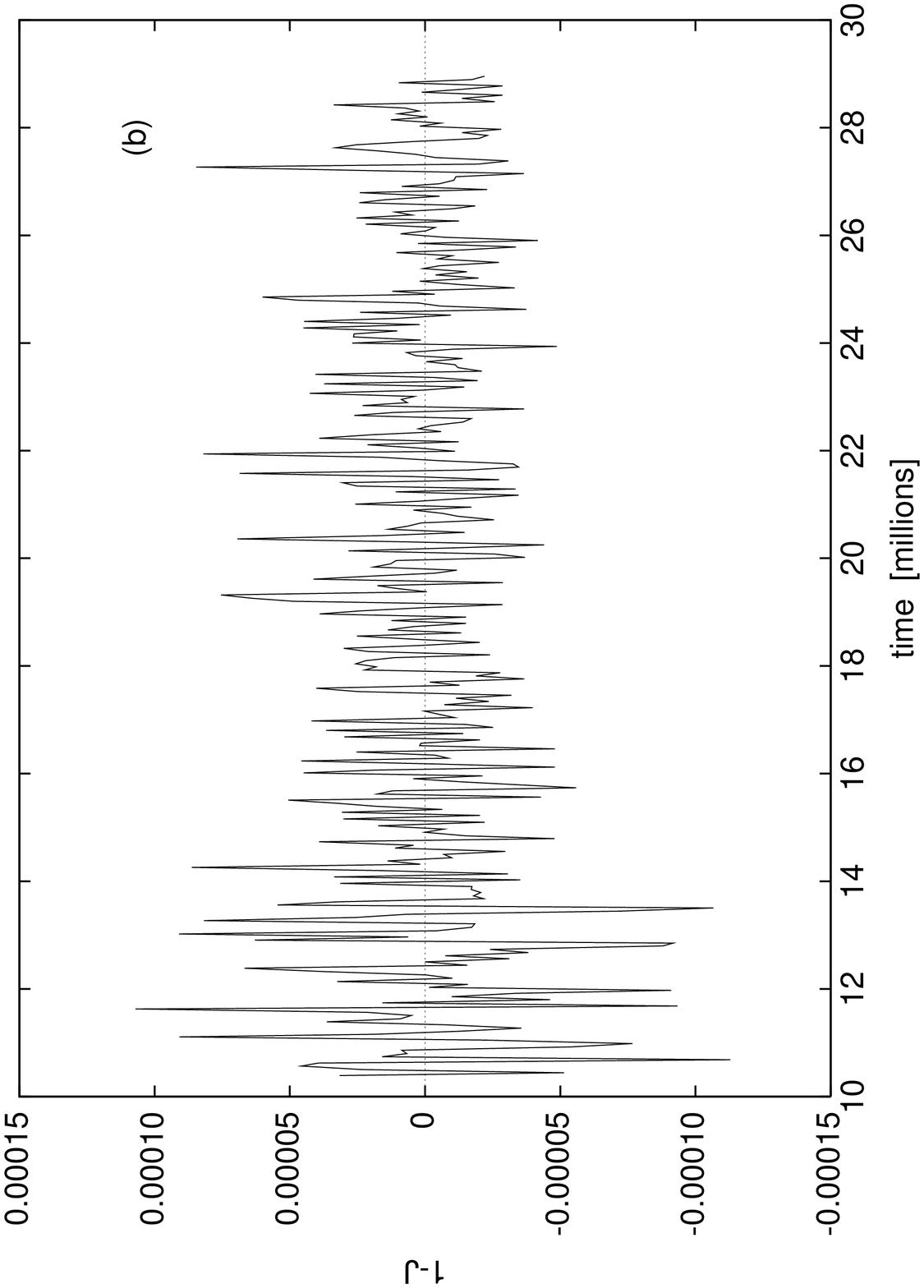}}}
}
\vspace*{0.5cm}
\caption{
Temporal dependence of the average absorbtion rate
(=``velocity'') for (a) $p=0.6$, and (b) $p=0.75$.
In both cases one observes fluctuations of about the same
order of magnitudes. However, although the system
is in the SOC state in both cases, the type of 
fluctuations are clearly different.
For $p>p_c$ one observes Gaussian white noise, whereas
$p<p_c$ leads to long-range correlations (reminiscent of
Brownian motion) in the transmitted signal.
For $p<p'$ everything gets absorbed, and there are no fluctuations.
\label{timeseq}}
\end{figure}

\begin{figure}
\narrowtext
\centerline{
	\epsfysize=\columnwidth{\rotate[r]{\epsfbox{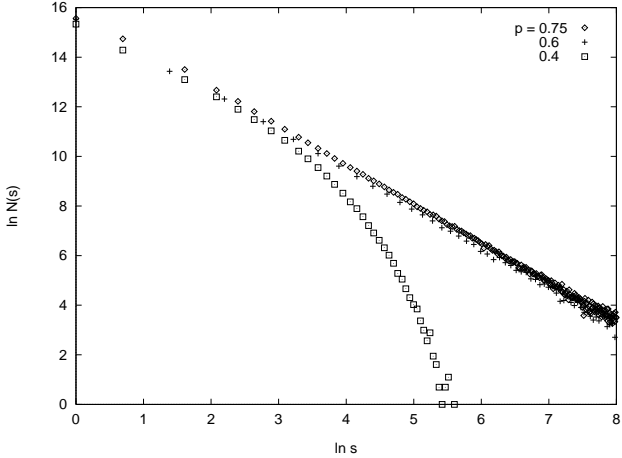}}}
}
\vspace*{0.5cm}
\caption{a)
Log-log plot of the avalanche size distributions for
$p=0.4$, $p=0.6$, and $p=0.75$.
The observed power-law scalings correspond to the
rice-pile universality class 
for $p>p'$, whereas the bounded avalanches for low $p$
show that $p<p'$ does not give critical fluctuations. 
\label{aval} }
\end{figure}

\begin{figure}
\narrowtext
\centerline{
	\epsfysize=\columnwidth{\rotate[r]{\epsfbox{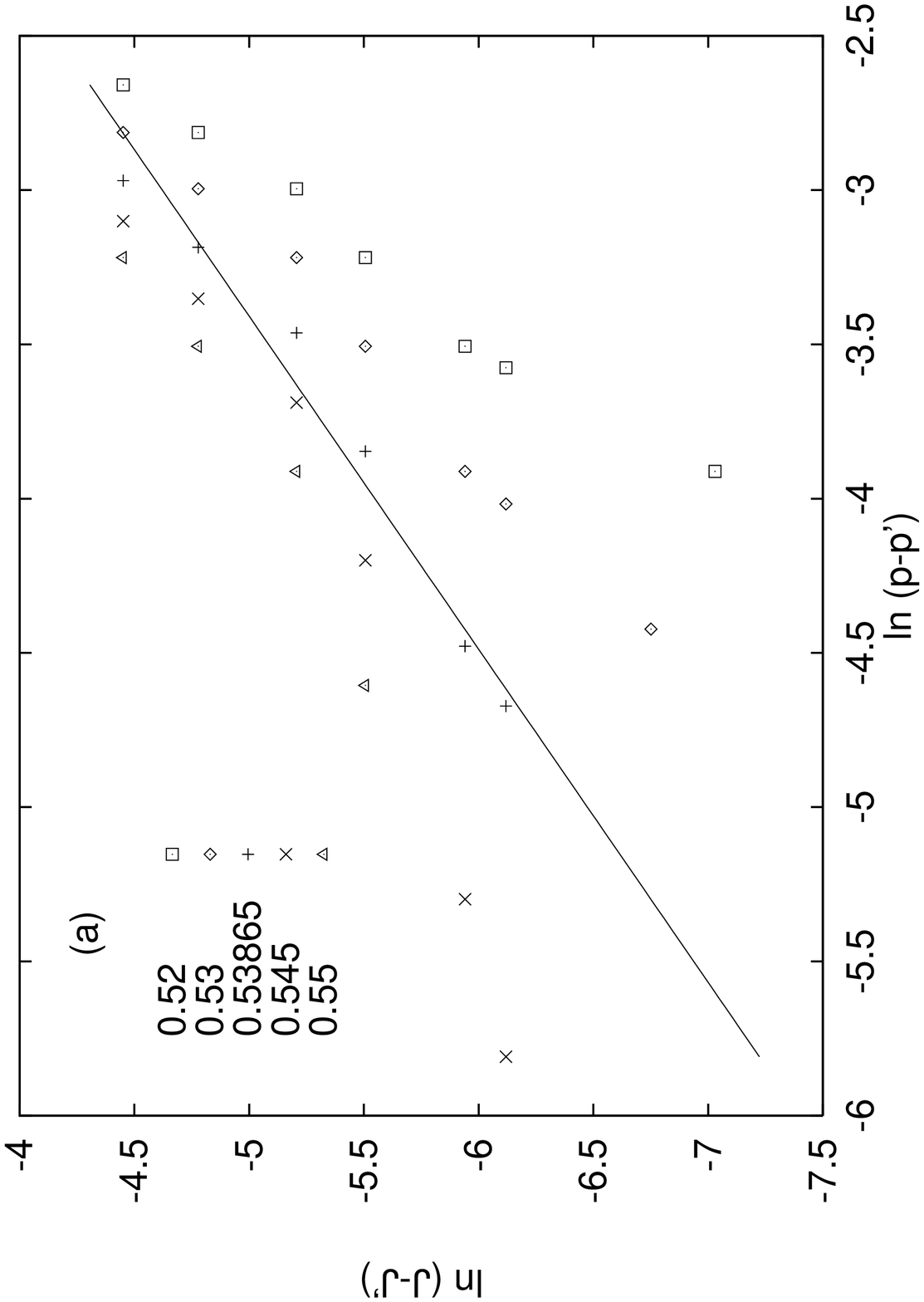}}}
}
\vspace*{0.5cm}
\centerline{
	\epsfysize=\columnwidth{\rotate[r]{\epsfbox{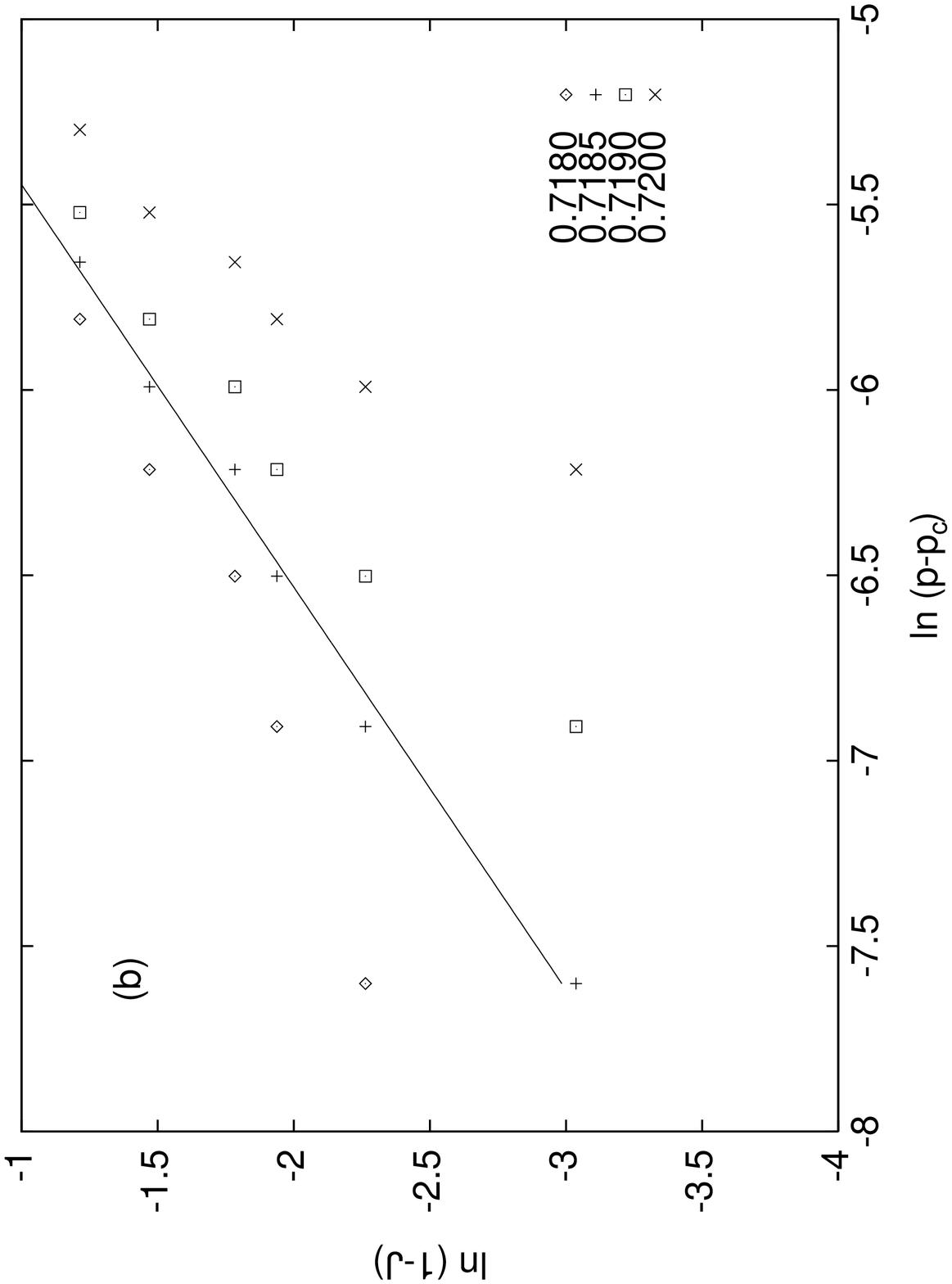}}}
}
\vspace*{0.5cm}
\caption{
Log-log plot of transport properties as a function
of distance from the critical points:
(a) The residual current $J(p)-J'$ vs.\ $p-p'$ for $p'$ values around
$0.53685$ (the legend shows the $p'$ values used).
Due to the finite system, there is a small
current $J' \equiv J(0.53685)$, and in order to improve the
scaling behavior we subtract $J'$ and find that
the best scaling is indeed obtained for $p'=0.53685$.
(b) The absorbtion (=$1-J$) for various $p_c$ values, cf.\
the legend. We find that the best scaling is obtained for $p_c=0.7185$.
\label{thres} }
\end{figure}

\end{multicols}


\begin{references}

\bibitem{NR}
A.M. Weinberg and E.P. Wigner, {\sl The Physical Theory of Nuclear
Chain Reactions}. (University of Chicago Press, Chicago, Illinois, 1958).
\bibitem{AD}
C.L. Hemenway, R.W. Henry, and M. Caulton, {\sl Physical
Electronics}, 2nd ed. (Wiley, New York, 1967).
\bibitem{LA}
H. Haken, {\sl Synergetics}, 3rd ed. (Springer, Heidelberg, 1983).
\bibitem{BTW}
P. Bak, C. Tang, and K. Wiesenfeld,
Phys. Rev. Lett. {\bf 59}, 381 (1987).
\bibitem{Zhang}
Y.-C. Zhang, Phys. Rev. Lett. {\bf 63}, 470 (1989).
\bibitem{Manna}
S.S. Manna, J. Phys. A {\bf 24}, L363 (1992).
\bibitem{Biham}
A. Ben-Hur and O. Biham,
Phys. Rev. E {\bf 53}, R1317 (1996).
\bibitem{CF}
K. Christensen, A. Corral, V. Frettte, J. Feder, and  T. J{\o}ssang,
Phys. Rev. Lett. {\bf 77}, 107 (1996).
\bibitem{Amaral96-1}
L.A.N. Amaral and K.B. Lauritsen,
Phys. Rev. E {\bf 54}, R4512 (1996);
Physica A {\bf 231}, 608 (1996).
\bibitem{Frette96}
V. Frette, K. Christensen, A. Malte-S{\o}rensen, J. Feder,
T. J{\o}ssang, and P. Meakin,
Nature {\bf 379}, 49 (1996).
\bibitem{PB}
M. Paczuski and S. Boettcher, Phys. Rev. Lett. {\bf 77}, 111 (1996).
\bibitem{NS}
H. Nakanishi and K. Sneppen, Preprint Nordita-96/49 S.
\bibitem{Kadanoff}
L.P. Kadanoff, S.R. Nagel, L. Wu, and S. Zhou,
Phys. Rev. A {\bf 39}, 6524 (1989).
\bibitem{Lana-Kbl}
L.A.N. Amaral and K.B. Lauritsen, preprint 1996.
\bibitem{Earth}
Z. Olami, H.J.S. Feder, and K. Christensen, 
Phys. Rev. Lett. {\bf 68}, 1244 (1992).

\end{references}
\end{document}